\documentclass[conference]{IEEEtran}
\IEEEoverridecommandlockouts
\usepackage{cite}
\usepackage{amsmath,amssymb,amsfonts}
\usepackage{algorithmic}
\usepackage{graphicx}
\usepackage{textcomp}
\usepackage{xcolor}

\usepackage[acronym]{glossaries}

\usepackage{booktabs}
\usepackage{svg}

\usepackage{tikz}
\usetikzlibrary{shapes, arrows}
\tikzset{>=latex}

\usepackage{setspace} 

\tikzstyle{block} = [draw, thick, rectangle, minimum height=0.75cm, minimum width=0.75cm]
\tikzstyle{sum} = [draw, fill=white, circle, node distance=1cm, thick]
\tikzstyle{gain} = [
  regular polygon,
  regular polygon sides=3,
  draw,
  thick,
  fill=white,
  text width=0.3em,
  inner sep=0.3mm,
  outer sep=0mm,
  shape border rotate=-90
]
\tikzstyle{revgain} = [
  regular polygon,
  regular polygon sides=3,
  regular polygon rotate=90,
  draw,
  thick,
  fill=white,
  text width=0.3em,
  inner sep=0.3mm,
  outer sep=0mm,
]

\usepackage{pgfplots}
\usepgfplotslibrary{fillbetween}
\usepgfplotslibrary{colormaps}

\usetikzlibrary{pgfplots.groupplots}

\definecolor{color0}{rgb}{0.12156862745098,0.466666666666667,0.705882352941177} 
\definecolor{color1}{rgb}{1,0.498039215686275,0.0549019607843137}
\definecolor{color2}{rgb}{0.172549019607843,0.627450980392157,0.172549019607843} 
\definecolor{color3}{rgb}{0.83921568627451,0.152941176470588,0.156862745098039} 
\definecolor{color4}{rgb}{0.580392156862745,0.403921568627451,0.741176470588235}
\pgfplotsset{
            compat=1.16,
            tick label style={font=\footnotesize},
            legend style={font=\footnotesize}
            }

\pgfplotscreateplotcyclelist{matplotlib}{
  {matplotlib0},
  {matplotlib1},
  {matplotlib2},
  {matplotlib3},
  {matplotlib4},
  {matplotlib5},
  {matplotlib6},
  {matplotlib7},
  {matplotlib8},
  {matplotlib9}
}

\pgfplotsset{every axis/.append style={
    cycle list name=matplotlib,
}}

\def\BibTeX{{\rm B\kern-.05em{\sc i\kern-.025em b}\kern-.08em
    T\kern-.1667em\lower.7ex\hbox{E}\kern-.125emX}}

\makeatletter
\def\ps@IEEEtitlepagestyle{%
  \def\@oddfoot{\mycopyrightnotice}%
  \def\@oddhead{\hbox{}\@IEEEheaderstyle\leftmark\hfil\thepage}\relax
  \def\@evenhead{\@IEEEheaderstyle\thepage\hfil\leftmark\hbox{}}\relax
  \def\@evenfoot{}%
}
\def\mycopyrightnotice{%
  \begin{minipage}{\textwidth}
  \centering \scriptsize
  \copyright 2023 IEEE.  Personal use of this material is permitted.  Permission from IEEE must be obtained for all other uses, in any current or future media, including reprinting/republishing this material for advertising or promotional purposes, creating new collective works, for resale or redistribution to servers or lists, or reuse of any copyrighted component of this work in other works.
  \end{minipage}
}
\makeatother
    
\begin{document}

\newcommand{\todo}[1]{{\color{red}#1}}
\newacronym{eeg}{EEG}{electroencephalography}
\newacronym{bmi}{BMI}{brain--machine interface}
\newacronym{snr}{SNR}{signal-to-noise ratio}
\newacronym{mm}{MM}{motor movement}
\newacronym{mi}{MI}{motor imagery}
\newacronym{cnn}{CNN}{convolutional neural network}
\newacronym{ble}{BLE}{Bluetooth low energy}
\newacronym{pcb}{PCB}{printed circuit board}
\newacronym{cmrr}{CMRR}{common-mode rejection ratio}
\newacronym{soa}{SoA}{state-of-the-art}
\newacronym{cv}{CV}{cross-validation}
\newacronym{tl}{TL}{transfer learning}
\newacronym{ulp}{ULP}{Ultra-Low Power}
\newacronym{afe}{AFE}{analog front-end}
\newacronym{pulp}{PULP}{parallel ultra-low power}
\newacronym{soc}{SoC}{system-on-chip}

\title{Enhancing Performance, Calibration Time and Efficiency in Brain-Machine Interfaces through Transfer Learning and Wearable EEG Technology
}

 \author{\IEEEauthorblockN{
    Xiaying Wang\IEEEauthorrefmark{1}\IEEEauthorrefmark{2},
    Lan Mei\IEEEauthorrefmark{2},
    Victor Kartsch\IEEEauthorrefmark{2}, 
    Andrea~Cossettini\IEEEauthorrefmark{2},
    Luca~Benini\IEEEauthorrefmark{2}\IEEEauthorrefmark{3}}
    \IEEEauthorblockA{\\\IEEEauthorrefmark{2}Dept. ITET, ETH Zurich, Zurich, Switzerland \hspace{3.2mm}\IEEEauthorrefmark{3}DEI, University of Bologna, Bologna, Italy\vspace{-0.2cm}}
    
    \thanks{* Corresponding email: xiaywang@iis.ee.ethz.ch}
    
    \vspace{-0.5cm}
    }

\maketitle

\begin{abstract}

Brain-Machine Interfaces (BMIs) have emerged as a transformative force in assistive technologies, empowering individuals with motor impairments by enabling device control and facilitating functional recovery. However, the persistent challenge of inter-session variability poses a significant hurdle, requiring time-consuming calibration at every new use. Compounding this issue, the low comfort level of current devices further restricts their usage. To address these challenges, we propose a comprehensive solution that combines a tiny CNN-based Transfer Learning (TL) approach with a comfortable, wearable EEG headband. The novel wearable EEG device features soft dry electrodes placed on the headband and is capable of on-board processing. We acquire multiple sessions of motor-movement EEG data and achieve up to 96\% inter-session accuracy using TL, greatly reducing the calibration time and improving usability. By executing the inference on the edge every 100\,ms, the system is estimated to achieve 30\,h of battery life. The comfortable BMI setup with tiny CNN and TL pave the way to future on-device continual learning, essential for tackling inter-session variability and improving usability. 

\end{abstract}


\section{Introduction}

\Gls{eeg} is the most common non-invasive technique used to measure and record the electrical activity of the brain. It is widely used in medical and research settings to study brain function, diagnose neurological disorders, and monitor the brain's status. Furthermore, EEG plays a central role in \glspl{bmi} bridging the human brain and external machines. One attractive \gls{bmi} paradigm is based on the motor function where the \gls{bmi} device can decode which body parts a subject moves---\gls{mm}---or imagines to move---\gls{mi}---based on \gls{eeg} signals~\cite{pfurtscheller2001functional}.


Over recent years, many \gls{eeg} devices have appeared on the consumer market. A few examples are Versus headset \cite{Versus}, Melomind \cite{Melomind}, Emotiv EPOC+ \cite{emotiv}, and Muse headband \cite{muse}. While they are much less intrusive and stigmatizing than clinically-used \gls{eeg} devices, they face a major problem related to low signal quality. Compared to the gold standard \gls{eeg} headsets based on gel, which are not suitable for out-of-lab daily usage mostly due to the long setup time and obtrusiveness, consumer-grade \gls{eeg} devices use dry electrodes to improve the user experience and reduce the setup time but yield low \gls{snr}. Authors in~\cite{Garcia2020_muse_headband_mi} used the Muse headband to collect \gls{mi} signals reaching a two-class classification accuracy ranging between 74\% and 93\%. However, a significant amount of data was polluted by noise, forcing the removal of up to 65\% of the acquired data.
%
Hence, it is extremely important to guarantee an optimal signal quality while improving the wearability of the \gls{eeg} devices to make them truly usable in daily life.

A second big challenge in \gls{eeg} is the inter-session (or intra-subject) variability: the signals acquired from the same person can vary when measured at different times (or sessions), despite using the same setup~\cite{saha2020_intersession}. As a result, the classification model trained on one session often yields significant accuracy degradation when applied to the next ones.

A simple solution is to train a new model for each session. However, this yields long calibration time, lowering user acceptance and engagement. 
Another way is to train big models on large datasets comprising multiple sessions. Even though there are various available open-source datasets, each uses different acquisition setups introducing inter-datasets variability, making the generalization problem even harder~\cite{Zaremba2022_cross-dataset}. Hence, it is a big challenge to directly use the models pre-trained on open-source datasets on commercially available devices. 
%
Another downside of this solution is that the models tend to grow in complexity and are impossible to be embedded on the edge device requiring the data to be transmitted to a remote gateway or cloud, yielding high power consumption and short battery lifetime. 
The latest trend of smart edge computing and tinyML has proven to be a promising solution for increasing the battery life of \gls{bmi} devices~\cite{wang2022mi,Kartsch2019_biowolf}. However, it requires a careful trade-off between accuracy performance and resource usage besides the necessity of miniaturized yet high-performing microprocessors.

In summary, the requirements for a \gls{bmi} device to be accepted by users and be usable in daily life scenarios are:
a) Compact, comfortable, and non-stigmatizing \gls{bmi} system; b) High classification accuracy with short calibration time; c) Energy-efficient design to prolong battery life and allow long-term usage.
%
In this paper, we present a comprehensive solution to fulfill the above requirements for an accurate and robust \gls{bmi} with low calibration time and high energy efficiency. 
The following points summarize our main contributions:
\begin{itemize}
    \item We propose a comfortable and non-stigmatizing \gls{eeg} headband featuring eight soft elastic elastomer-based dry active electrodes and BioWolf, a miniaturized acquisition device capable of onboard processing~\cite{Kartsch2019_biowolf}.
    \item We collect 7 sessions of \gls{eeg} data from one subject while performing left- and right-hand movements and address the inter-session variability using \gls{tl} with a tiny \gls{cnn}, i.e., MI-BMInet~\cite{wang2022mi}. We achieve up to 96\% inter-session accuracy, which is 30\% higher than the baseline without \gls{tl}.
    \item We also minimize the number of \gls{tl} training data and reduce by 2.13$\times$ the calibration time for data acquisition at every new session down to ca. 5 min. Additionally, the proposed "chain" \gls{tl} technique greatly reduces the resource requirements while performing \gls{tl}.
    \item MI-BMInet can be deployed on the \gls{pulp} RISC-V microprocessor on BioWolf taking circa 6\,ms and consuming up to 30\,uJ to execute one inference. 
    The system is estimated to provide more than 30\,h of battery life when doing inference every 100\,ms.
\end{itemize}





\section{Materials and Methods}

\subsection{EEG Headband}

\begin{figure}[t]
\centerline{\includegraphics[width=1\columnwidth]{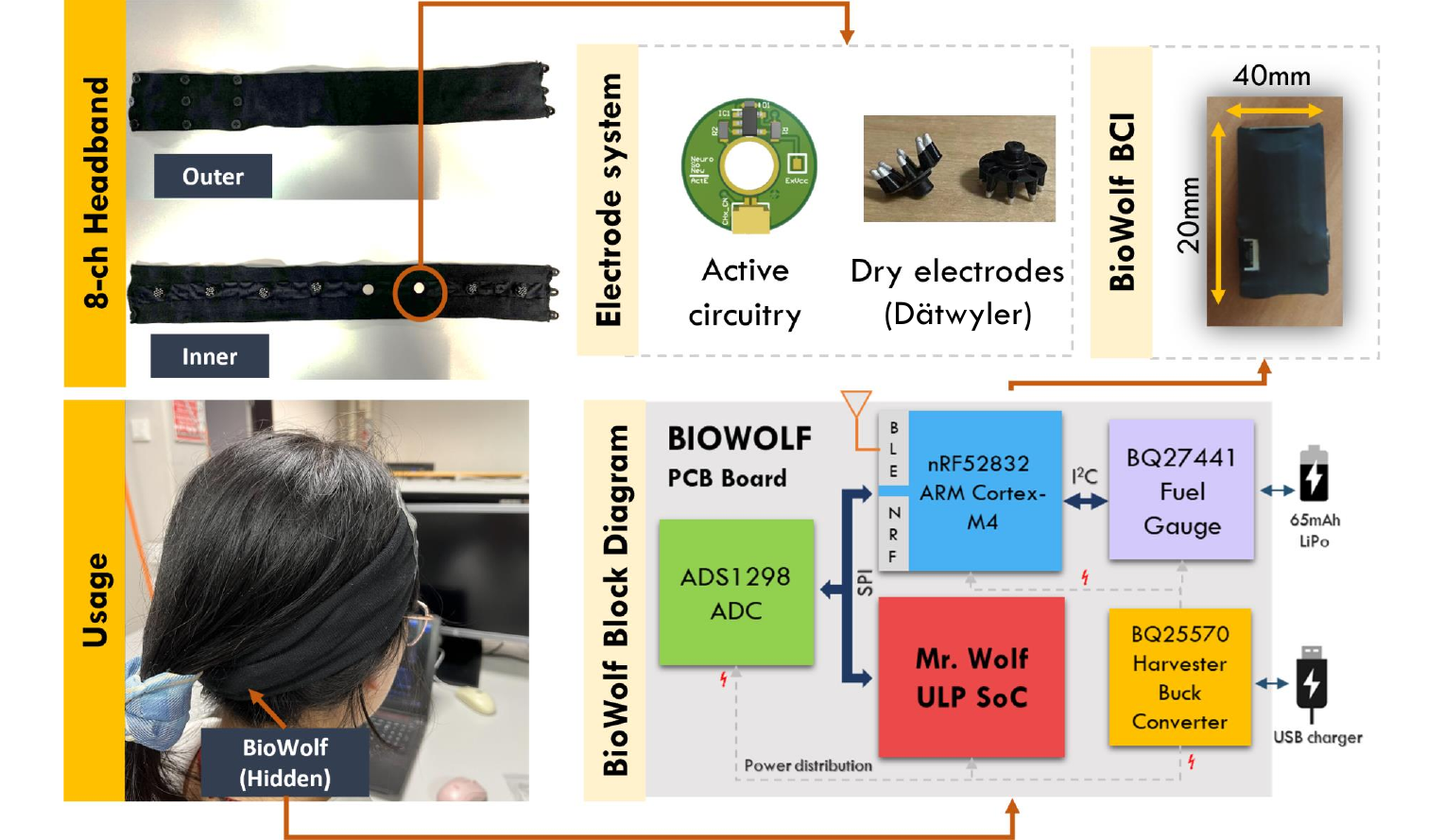}}
\vspace{-0.3cm}
\caption{Headband layout. Electrodes (D{\"a}twyler Schweiz AG) are integrated with an active buffering circuit, plugged into dedicated buttonholes on the headband and connected to BioWolf via cables hidden in the headband textile. 
}
\vspace{-0.2cm}
\label{fig:headband_nf}
\end{figure}

\Gls{eeg} data are collected with an ultra low-power \gls{bmi} platform (BioWolf~\cite{Kartsch2019_biowolf}), which features a \gls{pulp} \gls{soc} called Mr. Wolf, a 8-channel \gls{afe} for biopotentials \gls{eeg} channels, and a \gls{soc} with \gls{ble} for wireless connectivity.
BioWolf is integrated into a novel, comfortable, wearable and non-stigmatizing headband form-factor (Fig.~\ref{fig:headband_nf}). The in-house-tailored headband is made of comfortable clothing fabric, with a small pocket to hold the wearable \gls{bmi} device. The headband offers 3 different levels of tightness, which can be customized to the user's head size. 
Cables for connecting \gls{eeg} electrodes to BioWolf are hidden inside the headband (which appears as a normal sport headband that people can wear in daily life). The headband features 24 buttonholes for placing electrodes at different positions. For our experiments, we used 8 \gls{eeg} channels near the positions TP8, TP7, FC6, FC2, FC1, FC5, PO7, and PO8 of the standard 10/20 system. 

The electrode subsystem comprises two distinct mated components: the dry electrode and active buffering circuitry. The electrodes are fabricated by D{\"a}twyler Schweiz AG and are a new design part of the newly developed SoftPulse Flex family~\cite{DatwylerSoftPulse}. Electrodes are based on a soft elastic elastomer-based material and feature multiple silver/silver-chloride-coated legs arranged in a circular disc, enabling scalp contact and hair penetration without skin preparations or gel. An elastomer-based conductive snap connector allows to interface with the acquisition device.
The active circuitry, embedded in a circular \gls{pcb} and placed on top of the electrode, comprises a buffering subsystem, which enhances the \gls{cmrr} and reduces the artefacts caused by cable movement.

\subsection{Experimental Protocol}

A healthy volunteer is instructed to sit relaxed with the back against the chair and with the forearms lying on the lap. The subject follows visual instructions shown on a screen placed ca. 1\,m away from the subject's eyes, and performs \gls{mm} tasks of finger tapping~\cite{chang2018_bcitutorial}, while EEG signals are recorded using the wearable device at a sampling rate of 500\,Hz. Each instruction from the PC is synchronized with the \gls{eeg} acquisition using BioWolf digital triggering system.

\begin{figure}[t]
  \centering
  \includegraphics[width=0.9\linewidth, trim={7.5cm 9.6cm 9cm 3.65cm},clip]{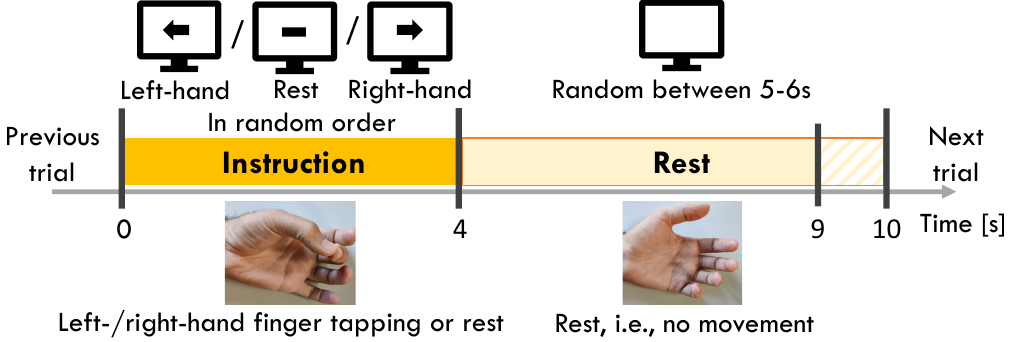}
  \vspace{-0.3cm}
  \caption{Timing scheme of one data collection trial.}
  \label{fig:trial_scheme}
\end{figure}

We collect seven sessions of \gls{eeg} data (a total of approx. 4.375 hours) from one subject, each session containing 12--20 runs.
One run consists of 15 trials, i.e., 5 with the left hand movement (left arrow), 5 with the right hand (right arrow), and 5 of resting state (dash line) in random order. As shown in Fig.~\ref{fig:trial_scheme}, a trial includes a 4\,s instruction and a resting period between 5\,s and 6\,s.
In this work, we perform two-class classfication of left- and right-hand movements.


\subsection{Classification Approach}

\subsubsection{Preprocessing}

The 8-channel \gls{eeg} data are preprocessed with a notch filter at 50\,Hz and a $4^{th}$ order IIR band-pass filter at 0.5--100\,Hz. We downsample the data in the 4\,s \gls{mm} window by a factor of 2 obtaining 950 samples in time. 

\begin{figure}[t]
  \centering
  \includegraphics[width=\linewidth, trim={0.83cm 3.3cm 1.25cm 3.5cm}, clip]{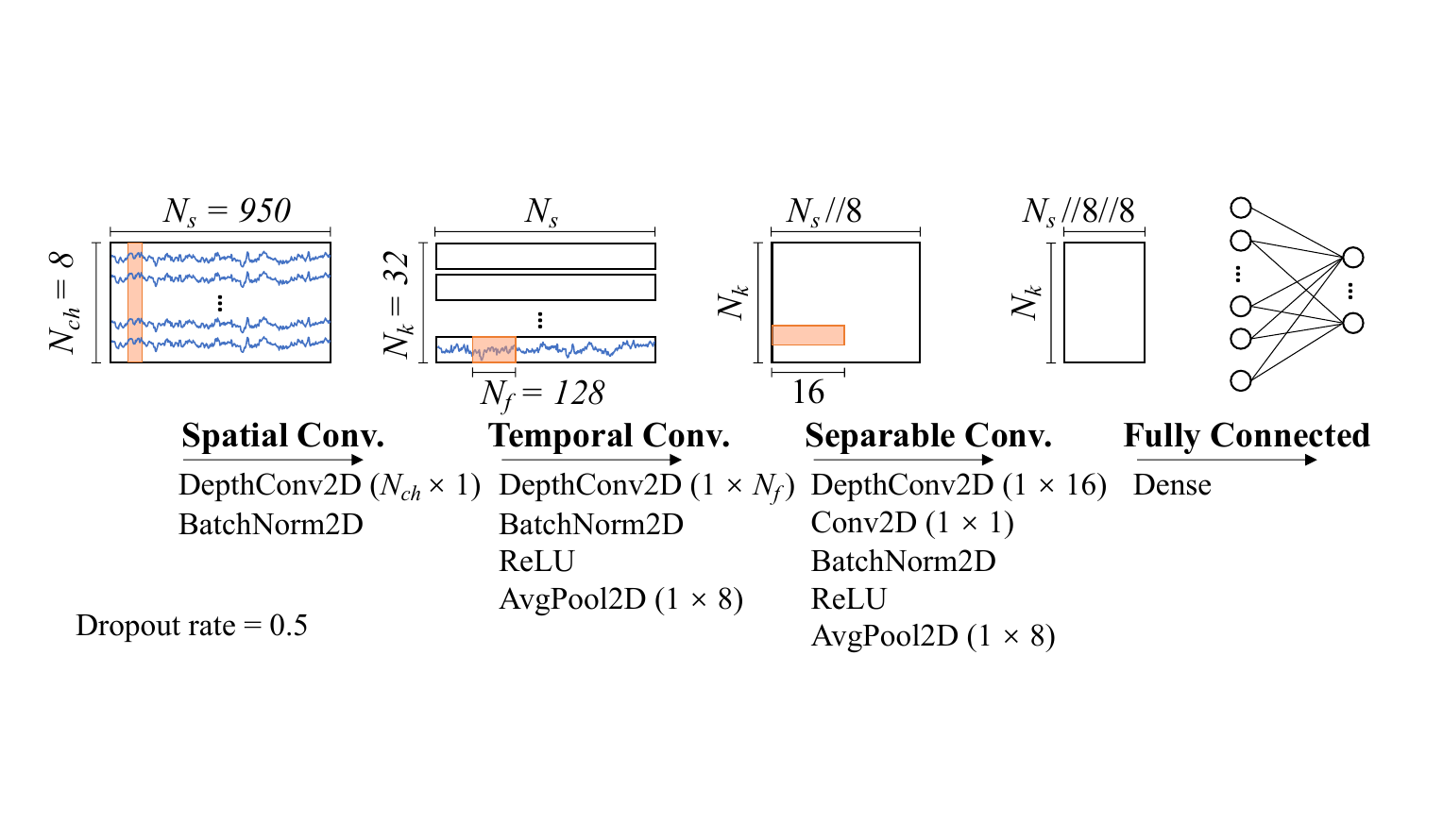}%
  \vspace{-0.2cm}
  \caption{MI-BMInet architecture~\cite{wang2022mi}.}
  \label{fig:MI-BMInet}
  \vspace{-0.2cm}
\end{figure}

\subsubsection{Within-session Training and Validation}


We use MI-BMInet~\cite{wang2022mi} which is a lightweight, embedded \gls{cnn} that achieves the \gls{soa} classification accuracy while demanding orders of magnitude less resources compared to other \gls{soa} models. The reduced resource usage is an essential requirement for embedded deployment on low-power edge devices.
%
Fig.~\ref{fig:MI-BMInet} shows the network architecture and the hyperparameters obtained after a grid search on 180 trials and 5-fold \gls{cv} experiments.

\begin{figure}[t]
  \centering
  \includegraphics[width=0.8\linewidth, trim={3cm 10cm 3cm 1cm}, clip]{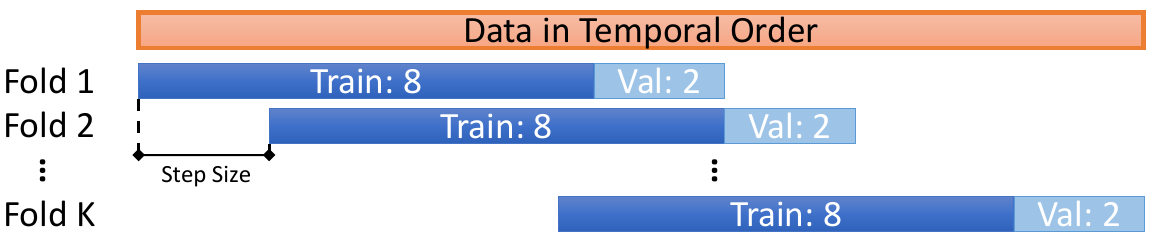}
  \vspace{-0.2cm}
  \caption{K-fold rolling \gls{cv}.}
  \vspace{-0.3cm}
  \label{fig:CVMethods}
\end{figure}

For the within-session baseline models, we use a rolling-window \gls{cv} approach, illustrated in Fig.~\ref{fig:CVMethods}, for training and validation. 
For each fold, the window of the training and validation data advances in time. This is to take into account the intrinsic changes of \gls{eeg} data in time. We also consider this method being the most suitable for future on-device learning as the model will need to be updated continuously with the latest data.
We train the models for 500 epochs using cross-entropy loss, Adam optimizer (lr=0.001), and a batch size of 64. In each fold of rolling-window \gls{cv}, eight runs, i.e., 80 trials were used as training set, and the subsequent two runs, i.e., 20 trials were used as validation set. The step size between neighboring folds is one run (Fig.~\ref{fig:CVMethods}). As in each session different number of runs were acquired, different number of folds in rolling-window \gls{cv} were applied. 


\subsection{Inter-session Transfer Learning}\label{subsec:intersessiontl}

To obtain the pre-trained model for \gls{tl}, we use the rolling-window \gls{cv} method as in the within-session case and choose the epoch $N_{ep}$ where the average validation loss and accuracy curves converge. We then train a model from scratch on all the available pretraining data until $N_{ep}$ and obtain a pre-trained model $M_{pre}$ that is ready to be used for \gls{tl} in the next session.

\begin{figure}[t]
  \centering
  \includegraphics[width=0.91\linewidth, trim={1cm 9cm 1.5cm 1cm}, clip]{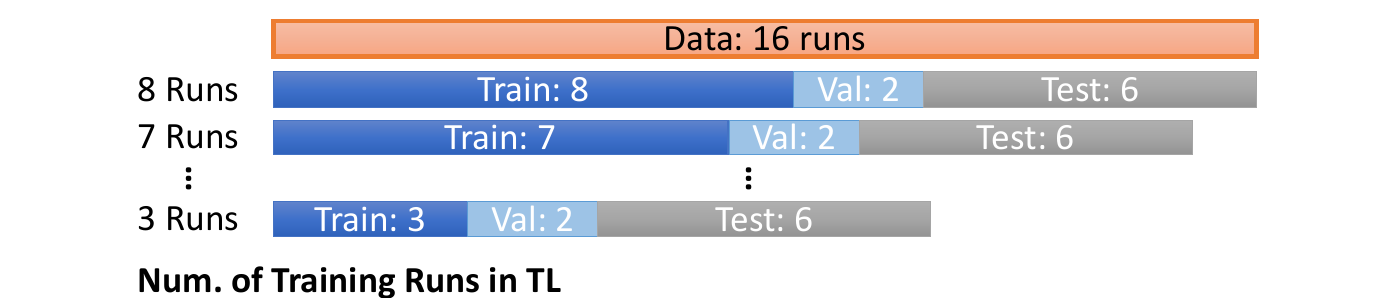}
  \vspace{-0.2cm}
  \caption{Train-validation-test split of \gls{tl} models with decreasing training runs.}
  \vspace{-0.3cm}
  \label{fig:NumTrainRuns}
\end{figure}

First, we use a one-to-one \gls{tl} approach: We load the pre-trained model $M_{pre}$ as the starting point to further train and validate on the first few runs of the new session. We select the final model $M_{TL}$ with the best \gls{tl} validation accuracy and test it on the remaining runs of this session in real time.
Fig.~\ref{fig:NumTrainRuns} illustrates an example with eight training runs, two validation runs, and six testing runs.
Additionally, we optimize the number of \gls{tl} training runs by performing the same procedure and reducing the number of training runs to assess the minimum number of new data necessary for \gls{tl} before the model is ready for real-time testing. This directly affects the calibration time.

Second, we explore a multi-to-one technique: Instead of using only one session of data for pre-training, we use multiple sessions to obtain the pre-trained model and perform the same \gls{tl} technique as before with the minimum number of \gls{tl} training runs on a new session.
With more pre-training sessions, the model is expected to learn more variability between sessions and be able to generalize better. 

A major problem of adding more sessions in the pre-training dataset is that every time a new session is recorded, a new pre-trained model has to be trained from scratch, requiring increasingly more memory for storing the dataset. Hence, we propose another experiment called ``chain" \gls{tl} scheme, where the new \gls{tl} model is constantly updated based on the previous \gls{tl} model.
Fig.~\ref{fig:ChainTL} shows an illustration: We obtain an initial $M_{pre}$ using one-session data and continuously perform \gls{tl} in the next sessions at every new usage of the headband by applying the same \gls{tl} technique with the minimum number of training runs followed by two validation runs.

\subsection{Drone Control}

We demonstrate the proposed methodology in a real use case scenario by conducting real-time testing and inference for drone control. The drone is a Crazyflie 2.1 from BitCraze~\cite{crazyflie} that can perform tasks while keeping a stable (level) flight autonomously. 
During inference, the drone receives the \gls{cnn} output commands (short left or right translations) using the BioWolf wireless link (BLE).

\section{Results}


\begin{table}[b]
\vspace{-0.3cm}
\renewcommand{\arraystretch}{0.9}
 \caption{Results of within-session rolling \gls{cv}. The average accuracy of all 7 sessions is 89.62\%.}
 \label{tab:WithinSession}
 \vspace{-0.1cm}
 \centering
 \resizebox{0.75\linewidth}{!}{
 \begin{tabular}{@{}llrr@{}} \toprule
  \textbf{Session} & \textbf{\#\,Runs} & \textbf{\#\,Folds} & \textbf{Avg. Accuracy $\pm$ Std. [\%]} \\ \midrule
  11.27 & 16 & 7  & $90.00 \pm 10.69$ \\
  11.30 & 16 & 7  & $86.43 \pm 10.25$ \\
  12.05 & 12 & 3  & $85.00 \pm 10.80$  \\
  12.09 & 13 & 4  & $81.25 \pm 2.17$ \\
  12.10 & 16 & 7  & $97.86 \pm 3.64$ \\
  12.13 & 12 & 3  & $95.00 \pm 4.08$ \\
  12.14 & 20 & 11  & $91.82 \pm 8.86$ \\ \bottomrule
 \end{tabular}
 }
\end{table}

\textbf{Within-session classification. } Table~\ref{tab:WithinSession} reports the best within-session classification \gls{cv} accuracy for each session.
Apart from Session 12.09, all the sessions' \gls{cv} accuracy exceeds 85\% with Session 12.10 reaching 97.86\%, which is comparable to \gls{soa} results for two-class classification~\cite{wang2022mi}. 
Note that in our experiments we perform motor movement instead of imagery since our focus is on the system level. However, previous studies show that finger movements and imagery have high real-virtual congruency, especially with four or five fingers movements, as in our case~\cite{diaz2008_real-virtual}.
If we directly compare to~\cite{Garcia2020_muse_headband_mi} in terms of average accuracy, ours is 6.12\% higher.


\textbf{One-to-one approach. }
We use Session 11.27 as pre-training data and Session 11.30 for \gls{tl} and real-time testing. 
With 7-fold rolling-window \gls{cv} on 11.27, we obtain $N_{ep}$=80 with 85\% \gls{cv} accuracy. We then obtain $M_{pre}$ by training a new model from scratch using all data from Session 11.27 for 80 epochs. 
Afterwards, \gls{tl} is applied on Session 11.30 as explained in Sec.~\ref{subsec:intersessiontl} and illustrated in Fig.\ref{fig:NumTrainRuns}. 
The model with the best \gls{tl} validation accuracy, $M_{TL}$, is finally selected and tested on the last 6 runs of the same session. To make a direct comparison without applying \gls{tl}, we test $M_{pre}$ on the same test runs and report the results in Table~\ref{tab:NumTrain}.
$M_{TL}$ showed a great improvement of 21.66\% in classification accuracy compared to $M_{pre}$, proving that the proposed \gls{tl} technique can successfully mitigate the inter-session variability and achieve high accuracy on newly-acquired data.

\textbf{Optimization of \gls{tl} training runs. }
We gradually reduce the number of \gls{tl} training runs from 8 to 3 to lower the calibration time in terms of acquiring new data (Fig.~\ref{fig:NumTrainRuns}).
%
The results are presented in Table~\ref{tab:NumTrain}. The \gls{tl} model using only three training runs can still achieve an excellent testing result of 96.67\%. Decreasing the number of training runs does not necessarily reduce the accuracy performance on the testing set, however, we can observe that more training epochs are necessary to obtain the model with the best validation accuracy. 
Meanwhile, by reducing the number of training runs from 8 to 3, we can reduce the new data acquisition time from about 17 minutes to about 8 minutes, i.e., 2.13$\times$ reduction, including the two validation runs. If we fix the number of epochs and eliminate the need of validation runs, the acquisition time is reduced down to ca. 5 minutes (i.e., 10 trials$\times$10\,s) with 90\% accuracy.

\begin{table}[b]
\vspace{-0.4cm}
\renewcommand{\arraystretch}{0.9}
 \caption{One-to-one inter-session classification accuracy w/ and w/o (marked by $^*$) \gls{tl} and with varying number of \gls{tl} training runs.
 }
 \label{tab:NumTrain}
 \vspace{-0.1cm}
 \centering
 \resizebox{0.85\linewidth}{!}{
 \begin{tabular}{@{}lrrrr@{}} \toprule
  \textbf{\# Runs} & \textbf{Best TL Epoch} & \textbf{Test Acc. [\%]} & \textbf{Acq. Time [min]} \\ \midrule
  8 & 121 & 88.33 & 16.7 \\
  - & - & 66.67$^*$  & - \\
  7 & 55 & 83.33  & 15 \\
  6 & 35 & 73.33  & 13.3 \\
  5 & 74 & 86.67  & 11.7 \\
  4 & 208 & 95.00  & 10 \\
  3 & 215 & 96.67 & 8.3 \\
  3 & 100 (fixed) & 90 & 5\\\bottomrule
 \end{tabular} 
 }
\end{table}


\textbf{Multi-to-one approach. }
First, we take 11.27 and 11.30 sessions as pre-training data and perform the same procedure as for one-to-one approach with a step size of 2 runs (Fig.~\ref{fig:CVMethods}). We obtain the pre-trained model $M_{pre2S}$ at $N_{ep}$ = 60 and perform \gls{tl} on Session 12.05 using the first 3 runs as \gls{tl} training and the following 2 runs as validation. The model $M_{TL2S}$ with the best validation accuracy is finally tested on the remaining 7 runs of Session 12.05 with results presented in Table~\ref{tab:NumTrainSessInd}. For comparison, we do the same testing with $M_{TL1S}$ which is obtained by applying \gls{tl} on Session 12.05 using the above $M_{pre}$. 
We can see that \gls{tl} offers an accuracy improvement of up to 31.43\% demonstrating once again the advantage of \gls{tl} techniques. When comparing $M_{TL2S}$ and $M_{TL1S}$, an improvement of 5.72\% is observed, inferring that more pre-training data might be beneficial for \gls{tl} to new sessions. 
Analogously, we conduct three-to-one \gls{tl} experiments on Session 12.13 
using session 11.27, 11.30, and 12.05 ($M_{TL3S}$) and the same ablation studies by reducing the number of pre-training sessions to two sessions 11.27 and 11.20 ($M_{TL2S}'$) and to one session 11.27 ($M_{TL1S}'$). For each experiment we provide also the comparison of test accuracy without applying \gls{tl}. 
The results are reported in Table~\ref{tab:NumTrainSessInd} and same conclusions can be drawn.





\begin{table}[b]
\vspace{-0.3cm}
\renewcommand{\arraystretch}{0.95}
 \caption{Multi-to-one inter-session test accuracy 
 w/ and w/o TL and comparison with lower number of pre-training sessions.}
 \label{tab:NumTrainSessInd}
 \vspace{-0.1cm}
 \centering
 \resizebox{0.75\linewidth}{!}{
 \begin{tabular}{@{}llrrr@{}} \toprule
  \textbf{Test on 12.05} &  \textbf{Test Acc.} & \textbf{Test on 12.13} &  \textbf{Test Acc.} &  \textbf{with TL?} \\ \midrule
    $M_{TL2S}$ & 81.43\% & $M_{TL3S}$ & 92.86\% & yes\\
  $M_{pre2S}$ & 50.00\% & $M_{pre3S}$ & 60.00\% & no \\
  $M_{TL1S}$ & 75.71\% & $M_{TL2S}'$ & 84.29\% & yes \\
  $M_{pre}$ & 50.00\% & $M_{pre2S}$ & 72.86\% & no\\
  - & - & $M_{TL1S}'$ & 81.43\% & yes \\
  - & - & $M_{pre}$ & 68.57\% & no\\
    \bottomrule
 \end{tabular} 
 }
\end{table}

\textbf{Chain \gls{tl}. } We pre-train an initial $M_{pre}$ model on Session 11.27. When we record the second session 11.30, we transfer $M_{pre}$ to $M_2$ using 3 training and 2 validation runs and test on the remaining runs. The same is done on a next session 12.09 by transferring $M_2$ to $M_3$ and so on.
We observe that chain \gls{tl} performs similarly well as the multi-to-one experiments and is a more promising solution for smart edge computing in continual learning. Every time a user uses a \gls{bmi} device in a new session, only a few recordings are necessary to calibrate the device and the \gls{tl} model update can be potentially also executed onboard, achieving a real-time update. 

\textbf{Embedded implementation. }
An inference of MI-BMInet deployed on Mr. Wolf takes approximately 6\,ms consuming up to 30\,uJ measured onboard~\cite{wang2022mi}. The power envelop of the whole system is estimated to be 8\,mW~\cite{wang2022mi,Kartsch2019_biowolf} while performing the inference every 100\,ms, which translates to 30\,h of operation with 65\,mAh battery. 

{
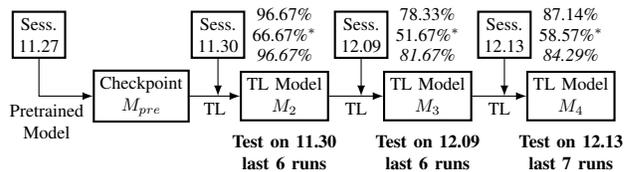
\begin{figure}[t]
  \begin{tikzpicture}[scale=0.68, transform shape, node distance=2cm]
    \node[block, align=center](Sess1){Sess. \\ 11.27};
    \node[block, right of=Sess1, xshift=0cm, yshift=-1.2cm, align=center](Mpre){Checkpoint \\ $M_{pre}$};
    \node[block, right of=Sess1, xshift=1.5cm, align=center](Sess2){Sess. \\  11.30};
    \node[block, right of=Sess2, xshift=0.8cm, align=center](Sess3){Sess. \\  12.09};
    \node[block, right of=Sess3, xshift=0.8cm, align=center](Sess4){Sess. \\  12.13};
    \node[block, right of=Mpre, xshift=0.8cm, align=center](M2){TL Model \\ $M_{2}$};
    \node[block, right of=M2, xshift=0.8cm, align=center](M3){TL Model \\ $M_{3}$};
    \node[block, right of=M3, xshift=0.8cm, align=center](M4){TL Model \\ $M_{4}$};
    \node[below of=M2, yshift=0.9cm, align=center](Test_1){\textbf{Test on 11.30} \\ \textbf{last 6 runs}};
    \node[below of=M3, yshift=0.9cm, align=center](Test_2){\textbf{Test on 12.09} \\ \textbf{last 6 runs}};
    \node[below of=M4, yshift=0.9cm, align=center](Test_3){\textbf{Test on 12.13} \\ \textbf{last 7 runs}};
    \node[above of=M2, yshift=-0.4cm, align=center](Acc_1){96.67\%};
    \node[above of=M2, yshift=-0.8cm, align=center](Acc_2){66.67\%$^*$};
    \node[above of=M2, yshift=-1.2cm, align=center](Acc_3){\textit{96.67\%}};
    \node[above of=M3, yshift=-0.4cm, align=center](Acc_4){78.33\%};
    \node[above of=M3, yshift=-0.8cm, align=center](Acc_5){51.67\%$^*$};
    \node[above of=M3, yshift=-1.2cm, align=center](Acc_6){\textit{81.67\%}};
    \node[above of=M4, yshift=-0.4cm, align=center](Acc_7){87.14\%};
    \node[above of=M4, yshift=-0.8cm, align=center](Acc_8){58.57\%$^*$};
    \node[above of=M4, yshift=-1.2cm, align=center](Acc_9){\textit{84.29\%}};

    \draw[->] (Sess1.south) |- (Mpre) node [left, xshift=-1cm, yshift=-0.5cm, align=center] {Pretrained \\ Model};
    \draw[->] (Mpre) -- node[below]{TL} (M2);
    \draw[->] (M2) -- node[below]{TL} (M3);
    \draw[->] (M3) -- node[below]{TL} (M4);
    \draw[->] (Sess2) -- + (0,-1.2);
    \draw[->] (Sess3) -- + (0,-1.2);
    \draw[->] (Sess4) -- + (0,-1.2);
  \end{tikzpicture}
  \vspace{-0.2cm}
  \caption{Chain \gls{tl} scheme with respective results and comparison without \gls{tl} (marked with $^*$) and with multi-to-one approach (italic).}\label{fig:ChainTL}
  \vspace{-0.3cm}
\end{figure}

}


\section{Conclusion}

We presented a novel solution for a comfortable, accurate, and energy-efficient \gls{bmi} based on \gls{eeg} signals and \gls{tl}. We designed and implemented a wearable \gls{eeg} headband with eight dry active electrodes and a miniaturized acquisition device that can perform onboard processing using a tiny \gls{cnn}. We collected and analyzed \gls{eeg} data from one subject performing left- and right-hand movements over seven sessions and demonstrated the effectiveness of \gls{tl} in overcoming inter-session variability.
We also deployed the inference of the tiny \gls{cnn} on a \gls{pulp} microprocessor to achieve low latency and long battery life.
%
For future works, we will investigate the resting class to account for cases of no commands, broaden our study to more subjects, and implement the \gls{tl} on the edge device for a continuous real-time adaptation of the \gls{bmi}.

\section*{Acknowledgment}

We thank Bettina Lory for implementing the initial experimental setup, Thorir Mar Ingolfsson for fruitful discussions, and D{\"a}twyler Schweiz AG for providing the electrodes.
This project is supported by the Swiss National Science Foundation under the grant number 193813 (PEDESITE project) and the grant number 207913 (TinyTrainer project) and by the ETH Future Computing Laboratory (EFCL).

\bibliographystyle{IEEEtran}
\bibliography{bib}

\begin{thebibliography}{10}
\providecommand{\url}[1]{#1}
\csname url@samestyle\endcsname
\providecommand{\newblock}{\relax}
\providecommand{\bibinfo}[2]{#2}
\providecommand{\BIBentrySTDinterwordspacing}{\spaceskip=0pt\relax}
\providecommand{\BIBentryALTinterwordstretchfactor}{4}
\providecommand{\BIBentryALTinterwordspacing}{\spaceskip=\fontdimen2\font plus
\BIBentryALTinterwordstretchfactor\fontdimen3\font minus
  \fontdimen4\font\relax}
\providecommand{\BIBforeignlanguage}[2]{{%
\expandafter\ifx\csname l@#1\endcsname\relax
\typeout{** WARNING: IEEEtran.bst: No hyphenation pattern has been}%
\typeout{** loaded for the language `#1'. Using the pattern for}%
\typeout{** the default language instead.}%
\else
\language=\csname l@#1\endcsname
\fi
#2}}
\providecommand{\BIBdecl}{\relax}
\BIBdecl

\bibitem{pfurtscheller2001functional}
G.~Pfurtscheller, ``Functional brain imaging based on erd/ers,'' \emph{Vision
  research}, vol.~41, no. 10-11, pp. 1257--1260, 2001.

\bibitem{Versus}
N.~M. LLC, ``Versus: a mobile {EEG} headset,'' \url{https://getversus.com/},
  2021, accessed: 2021-10-11.

\bibitem{Melomind}
G.~Spinelli, A.~Odouard, M.-C. Nierat, S.~Campion, M.~Bensoussan, F.~Grosselin,
  K.~Pandremmenou, A.~Breton, M.~Raux, Y.~Attal, T.~Similowski, and
  X.~Navarro-Sun{\'e}, ``Validation of melomind{\texttrademark} signal quality:
  a proof of concept resting-state and {ERPs} study,''
  \emph{bioRxiv:2020.02.28.969808}, 2020.

\bibitem{emotiv}
{Emotiv}, ``Epoc+,'' \url{https://www.emotiv.com/epoc/}, accessed: 2023-05-30.

\bibitem{muse}
{Muse}, ``{EEG-powered meditation and sleep headband},''
  \url{https://choosemuse.com/}, accessed: 2023-05-30.

\bibitem{Garcia2020_muse_headband_mi}
\BIBentryALTinterwordspacing
F.~M. Garcia-Moreno, M.~Bermudez-Edo, J.~L. Garrido, and M.~J.
  Rodríguez-Fórtiz, ``Reducing response time in motor imagery using a
  headband and deep learning,'' \emph{Sensors}, vol.~20, no.~23, 2020.
  [Online]. Available: \url{https://www.mdpi.com/1424-8220/20/23/6730}
\BIBentrySTDinterwordspacing

\bibitem{saha2020_intersession}
\BIBentryALTinterwordspacing
S.~Saha and M.~Baumert, ``Intra- and inter-subject variability in eeg-based
  sensorimotor brain computer interface: A review,'' \emph{Frontiers in
  Computational Neuroscience}, vol.~13, 2020. [Online]. Available:
  \url{https://www.frontiersin.org/articles/10.3389/fncom.2019.00087}
\BIBentrySTDinterwordspacing

\bibitem{Zaremba2022_cross-dataset}
T.~Zaremba and A.~Atyabi, ``Cross-subject \& cross-dataset subject transfer in
  motor imagery bci systems,'' in \emph{2022 International Joint Conference on
  Neural Networks (IJCNN)}, 2022, pp. 1--8.

\bibitem{wang2022mi}
X.~Wang, M.~Hersche, M.~Magno, and L.~Benini, ``Mi-bminet: An efficient
  convolutional neural network for motor imagery brain--machine interfaces with
  eeg channel selection,'' \emph{arXiv preprint arXiv:2203.14592}, 2022.

\bibitem{Kartsch2019_biowolf}
V.~Kartsch, G.~Tagliavini, M.~Guermandi, S.~Benatti, D.~Rossi, and L.~Benini,
  ``Biowolf: A sub-10-mw 8-channel advanced brain–computer interface platform
  with a nine-core processor and ble connectivity,'' \emph{IEEE Transactions on
  Biomedical Circuits and Systems}, vol.~13, no.~5, pp. 893--906, 2019.

\bibitem{DatwylerSoftPulse}
{Datwyler Schweiz AG}, ``Softpulse,'' \url{https://datwyler.com/company/
  innovation/wearable-sensors/softpulse}.

\bibitem{chang2018_bcitutorial}
H.~Cho, M.~Ahn, M.~Kwon, and S.~Jun, ``A step-by-step tutorial for a motor
  imagery–based {BCI},'' in \emph{Brain–Computer Interfaces Handbook:
  Technological and Theoretical Advances}, F.~L. Chang S.~Nam, Anton~Nijholt,
  Ed.\hskip 1em plus 0.5em minus 0.4em\relax Taylor \& Francis Group, 2018,
  ch.~23, pp. 445--460.

\bibitem{crazyflie}
\BIBentryALTinterwordspacing
{BitCraze}, ``{Crazyflie 2.1}.'' [Online]. Available:
  \url{https://www.bitcraze.io/about/bitcraze/}
\BIBentrySTDinterwordspacing

\bibitem{diaz2008_real-virtual}
M.~Díaz, C.~Llanos, S.~Gonzalez, and M.~Sabate, ``How similar are motor
  imagery and movement?'' \emph{Behavioral neuroscience}, vol. 122, pp. 910--6,
  09 2008.

\end{thebibliography}

\end{document}